# The Dominant Role of Critical Valence Fluctuations on High $T_\mathrm{c}$ Superconductivity in Heavy Fermions

Gernot W. Scheerer*

*DQMP - University of Geneva, 1211 Geneva 4, Switzerland.*

Zhi Ren

*Institute of Natural Sciences, Westlake Institute For Advanced Study,*
*Westlake University, 18 Shilongshan Road, Hangzhou 310024, P. R. China.*

Shinji Watanabe

*Department of Basic Sciences, Kyushu Institute of Technology,*
*Kitakyushu, Fukuoka, 804-8550, Japan.*

Gérard Lapertot

*PHELIQS, UMR-E 9001, CEA-INAC/UJF-Grenoble 1, 38054 Grenoble, France.*

Dai Aoki

*Institute for Materials Research, Tohoku University,*
*Oarai, Ibaraki, 311-1313, Japan and*
*PHELIQS, UMR-E 9001, CEA-INAC/UJF-Grenoble 1, 38054 Grenoble, France.*

Didier Jaccard

*DQMP - University of Geneva, 1211 Geneva 4, Switzerland.*

Kazumasa Miyake

*Center for Advanced High Magnetic Field Science,*
*Osaka University, Toyonaka, Osaka, 560-0043, Japan.*

## Abstract

Despite almost 40 years of research, the origin of heavy-fermion superconductivity is still strongly debated. Especially, the pressure-induced enhancement of superconductivity in $CeCu_2Si_2$ away from the magnetic breakdown is not sufficiently taken into consideration. As recently reported in $CeCu_2Si_2$ and several related compounds, optimal superconductivity occurs at the pressure of a valence crossover, which arises from a virtual critical end point at negative temperature $T_{cr}$. In this context, we did a meticulous analysis of a vast set of top-quality high-pressure electrical resistivity data of several Ce-based heavy fermion compounds. The key novelty is the salient correlation between the superconducting transition temperature $T_c$ and the valence instability parameter $T_{cr}$, which is in line with theory of enhanced valence fluctuations. Moreover, it is found that, in the pressure region of superconductivity, electrical resistivity is governed by the valence crossover, which most often manifests in scaling behavior. We develop the new idea that the optimum superconducting $T_c$ of a given sample is mainly controlled by the compound's $T_{cr}$ and limited by non-magnetic disorder. In this regard, the present study provides compelling evidence for the crucial role of critical valence fluctuations in the formation of Cooper pairs in Ce-based heavy fermion superconductors besides the contribution of spin fluctuations near magnetic quantum critical points, and corroborates a plausible superconducting mechanism in strongly correlated electron systems in general.

* gernot.scheerer@unige.ch

# I. INTRODUCTION

Superconductivity (SC) in heavy fermion (HF) systems is most often considered as being mediated by critical spin fluctuations [1–4]. Such a prevailing view is mainly derived from the presence of a magnetic instability regime leading to the collapse of long-range antiferromagnetic (AF) order at a critical $p_c$, concomitant with the emergence of SC. However in a few cases, SC has been ascribed to critical valence fluctuations (CVF) in the pressure region of the highest superconducting transition temperature $T_c$, in particular for $CeCu_2Ge_2$ [5, 6], $CeCu_2Si_2$ [7, 8], and $CeRhIn_5$ [9, 10]. The main ingredient of this interpretation is the existence in the pressure-temperature ($p$-$T$) plane of an underlying first-order valence transition, whose critical end point (CEP) occurs at pressure $p_{cr}$ and at slightly negative temperature $T_{cr}$ (see Fig. 1). With a negative $T_{cr}$, only a valence crossover (VCO) regime is accessible at finite temperature and the corresponding crossover line lies close to optimal SC. In the case of the prototype HF superconductor $CeCu_2Si_2$, multiple experimental evidence of the VCO and CVF mediated SC has been reported in [5, 7, 8, 11–14]. For instance, direct, microscopic observation of the VCO and the absence of spin fluctuations close to optimal SC have been reported for $CeCu_2Si_2$ [13] and also $CeIrIn_5$ [15, 16] via Cu- and In-nuclear quadrupole resonance measurements, respectively.

Selected examples of $p$-$T$ magnetic and superconducting phase diagrams of Ce-based HF superconductors are represented schematically in Fig. 1. The common feature of all compounds is that SC is optimal at a pressure close to $p_{cr}$. On the other hand, the magnetic $p_c$ can coincide with $p_{cr}$ as in the case of $CeRhIn_5$ [17], $CePd_2Si_2$ (this work), and $CeAu_2Si_2$ [18], or be much lower than $p_{cr}$ as in $CeCu_2Si_2$ [8]. The spreading of SC over the pressure axis varies considerably and SC can even emerge deep inside the magnetic phase of $CeAu_2Si_2$ [18].

As a matter of interest, the CVF mechanism shares common aspects with the d-p charge transfer instability in high-$T_c$ cuprates, which has been proposed to be at the origin of marginal Fermi liquid and non-Fermi liquid properties, and the pseudo-gap state [20, 21]. Moreover, valence fluctuations of Pu ions have been advocated as the source of "high-$T_c$" in $PuCoGa_5$ [22]. Thus, we believe that the valence fluctuation physics discussed in this paper is pertinent for a larger community beyond that of HFs.

The microscopic-theoretical basis of the CVF scenario results from the inclusion of the additional term $H_{U_{fc}} = U_{fc} \sum_{i=1}^{N} n_i^f n_i^c$ in the periodic Anderson model, where $U_{fc}$ is the Coulomb repulsion between f and conduction electrons [23, 24]. The physical origin of the $T_c$ enhancement is the increase in the effective quasiparticle Fermi energy and the constancy of the dimensionless coupling for the Cooper pairing, following a BCS-like expression for $T_c$. The former factor stems from the valence crossover from the Kondo to the valence fluctuation region and the latter one is a result of the compensation between the decrease of the quasiparticle density of states and the increase in the pairing interaction, which is mediated by valence fluctuations associated with a sharp valence crossover.

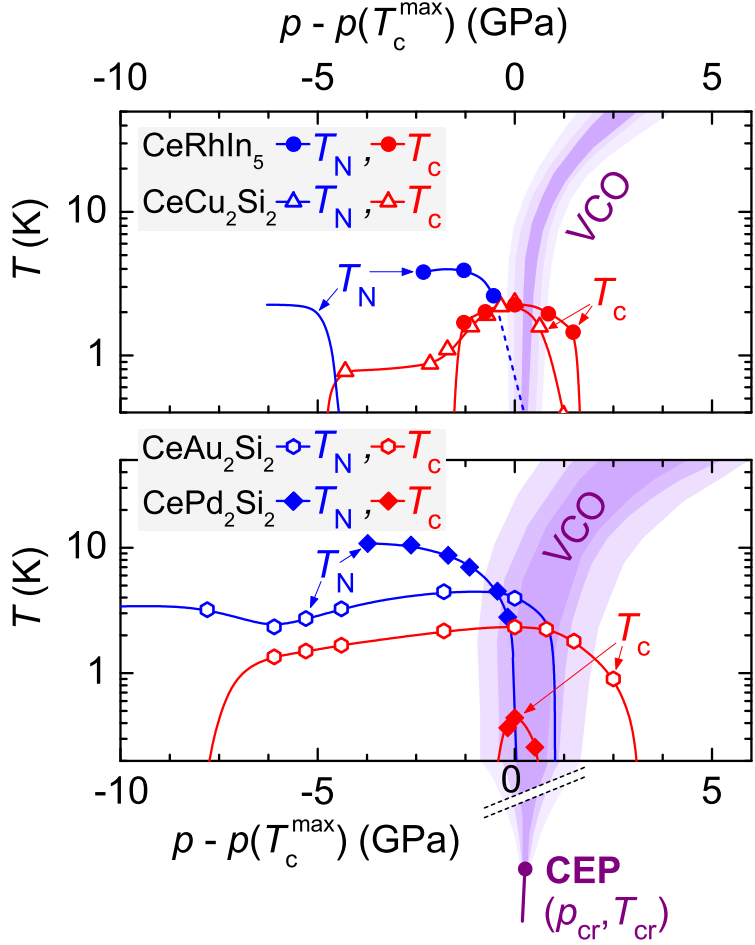

FIG. 1. **Examples of schematic $p$-$T$ magnetic and superconducting phase diagrams of Ce-based HF superconductors.** Symbols stand for representative data points from CeRhIn$_5$ [17], CeCu$_2$Si$_2$ [8], CeAu$_2$Si$_2$ [18], and CePd$_2$Si$_2$ [19]: Néel temperature $T_N$ and bulk-superconducting $T_c$. The graded-colored area represents the valence crossover.

Five years ago, thanks to an experimental progress [25] yielding more accurate electrical resistivity measurements on CeCu$_2$Si$_2$ under pressure up to 7 GPa, we have introduced a method to estimate the temperature $T_{\mathrm{cr}}$ of the CEP [8]. Subsequently, the same process was successfully used for $CeAu_2Si_2$ [18, 26, 27]. In the present paper, this method is applied to all appropriate resistivity data established in Geneva since 1998, including new data notably from $CeAg_2Si_2$, $CeRhIn_5$ and $CeIrIn_5$. On the basis of 17 data sets from 9 different Ce-based HF compounds, the universal character of the relationship between the superconducting transition temperature $T_{\mathrm{c}}$ and the strength of the valence instability is unveiled. Taking into account the superconducting pair-breaking effect of non-magnetic disorder, quantified by the residual resistivity $\rho_0$, we identify the two main parameters $T_{\mathrm{cr}}$ and $\rho_0$ controlling $T_{\mathrm{c}}$ of a representative part of Ce-based HF superconductors, which is consistent with the CVF theory. Moreover, it is found that, in the VCO regime of the $p$-$T$ plane, electrical resistivity most often follows scaling behavior, underlining the role of valence fluctuation physics.

## II. RESULTS

Figure 2 displays a 3D plot of the superconducting $T_{\mathrm{c}}$ as function of both the residual resistivity $\rho_0$ and the valence instability parameter $T_{\mathrm{cr}}$ based on published and new results (see Table S1 of the Supplementary Material for details and references). In this paper "$T_{\mathrm{c}}$" refers to the maximum value of the bulk-superconducting transition temperature versus pressure for a given sample. Evidently in Fig. 2, all compounds except $CeCu_5Au$ lie more or less on an empirically drawn inclined surface with a maximum for small $\rho_0$ and $T_{\mathrm{cr}}$, which suggests that the superconducting $T_{\mathrm{c}}$ of a given sample is mainly controlled by the compound's $T_{\mathrm{cr}}$ and the sample's $\rho_0$. $T_{\mathrm{c}}$ seems to culminate at $\sim 2.5$ K when $T_{\mathrm{cr}} \to 0$ and $\rho_0 \to 0$, i.e., for a quantum CEP and negligible pair breaking effect. However, high $\rho_0$ values or large negative $T_{\mathrm{cr}}$ depress $T_{\mathrm{c}}$.

We underline that all samples with $T_{\mathrm{c}} > 2$ K are found to exhibit -15 K $< T_{\mathrm{cr}} < 0$ K and emphasize the striking relationship between the superconducting $T_{\mathrm{c}}$, the parameter $T_{\mathrm{cr}}$ of the valence transition CEP, and pair breaking due to non-magnetic impurities ($\rho_0$). We introduce the expression "high-$T_{\mathrm{c}}$" to refer to the fact that the compounds with the highest $T_{\mathrm{c}}$ amongst the Ce-based HF superconductors are especially well represented in Fig. 2. Five out of the nine studied compounds have $T_{\mathrm{c}}$ higher than 1.5 K. At the moment, important cases like $CeCoIn_5$ [28], $CeRh_2Si_2$ [29], $CePt_2In_7$ [30], or non-centrosymmetric $CePt_3Si$ [31] are lacking for different reasons (see below). Nevertheless, Fig. 2 represents a substantial

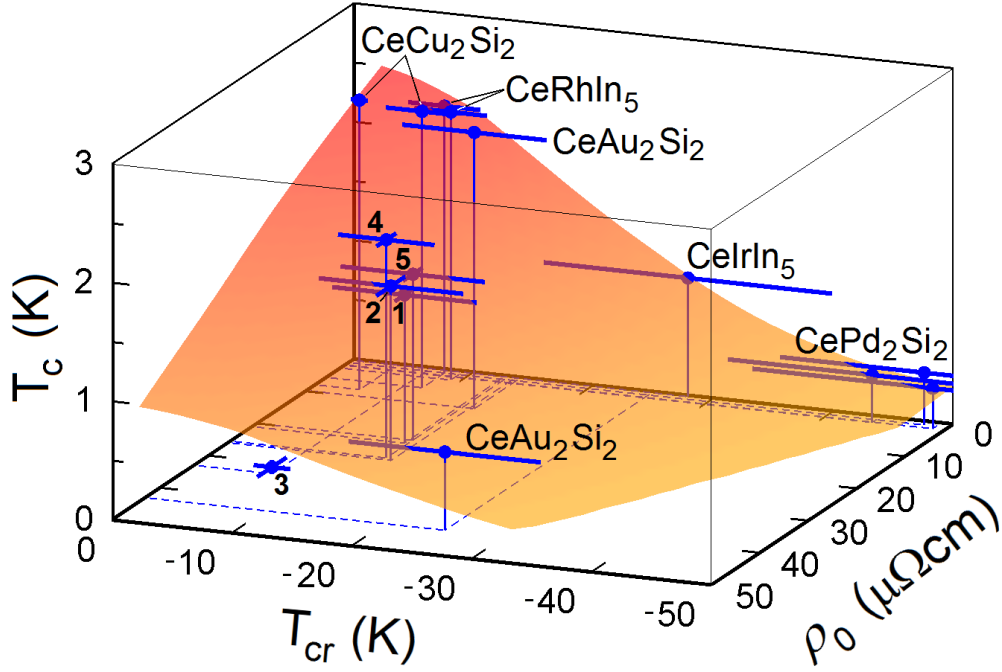

FIG. 2. **Maximal superconducting $T_c$ of Ce-based HF superconductors as a function of the key parameter $T_{cr}$ and the residual resistivity $\rho_0$.** See text for details. Bold numbers indicate: CeAg$_2$Si$_2$ = 1, CeCu$_2$Ge$_2$ = 2 , CeCu$_5$Au = 3, and other samples of CeCu$_2$Si$_2$ = 4 and CeAu$_2$Si$_2$ = 5. The error bars for $T_{cr}$ represent estimated errors according the scaling analysis (see main text) and the error bars for $\rho_0$ result from the power-law extrapolation to zero temperature of $\rho(T)$. All compounds lie on or not far from the empirically drawn surface. Blue (violet) data points lie above (below) the surface. The surface is drawn for $T_{cr} \leq -2$ K for a reason discussed below and for $T_c \geq 0.28$ K, since no reliable information exists for very negative $T_{cr}$ or very high $\rho_0$.

part of Ce-based HF compounds and gives a unified view on their SC.

Before taking a closer look to the relationships $T_c(T_{cr})$ and $T_c(\rho_0)$, let us discuss the behavior of electrical resistivity $\rho$ in the VCO regime and summarize the method for extracting $T_{cr}$ from low-temperature $\rho$ [8]. First, in Figure 3(a) we compare the schematic $p$-$T$ phase diagrams of CeCu$_2$Si$_2$ and elementary Ce. In Ce, a first-order valence transition (FOVT) occurs at finite temperature due to small Ce-Ce ion spacing and therefore strong $U_{fc}$-repulsion between f- and conduction electrons at the same Ce site. The CEP lies at $p_{cr} \approx 1.5$ GPa and $T_{cr} \approx 480$ K [32]. As a function of pressure, isothermal resistivity of Ce [Fig. 3(b)] exhibits a discontinuous anomaly at the FOVT ($T < T_{cr}$) [33]. In the crossover regime ($T > T_{cr}$), isothermal resistivity decreases rapidly but continuously and the resistivity gradient diverges just at the CEP ($T \to T_{cr}$). In CeCu$_2$Si$_2$, the CEP lies at slightly negative temperature $T_{cr} \approx -8$ K and, in the VCO regime, isothermal $\rho$ decreases more and more rapidly versus pressure as temperature goes down without reaching a first-order discontinuity [8].

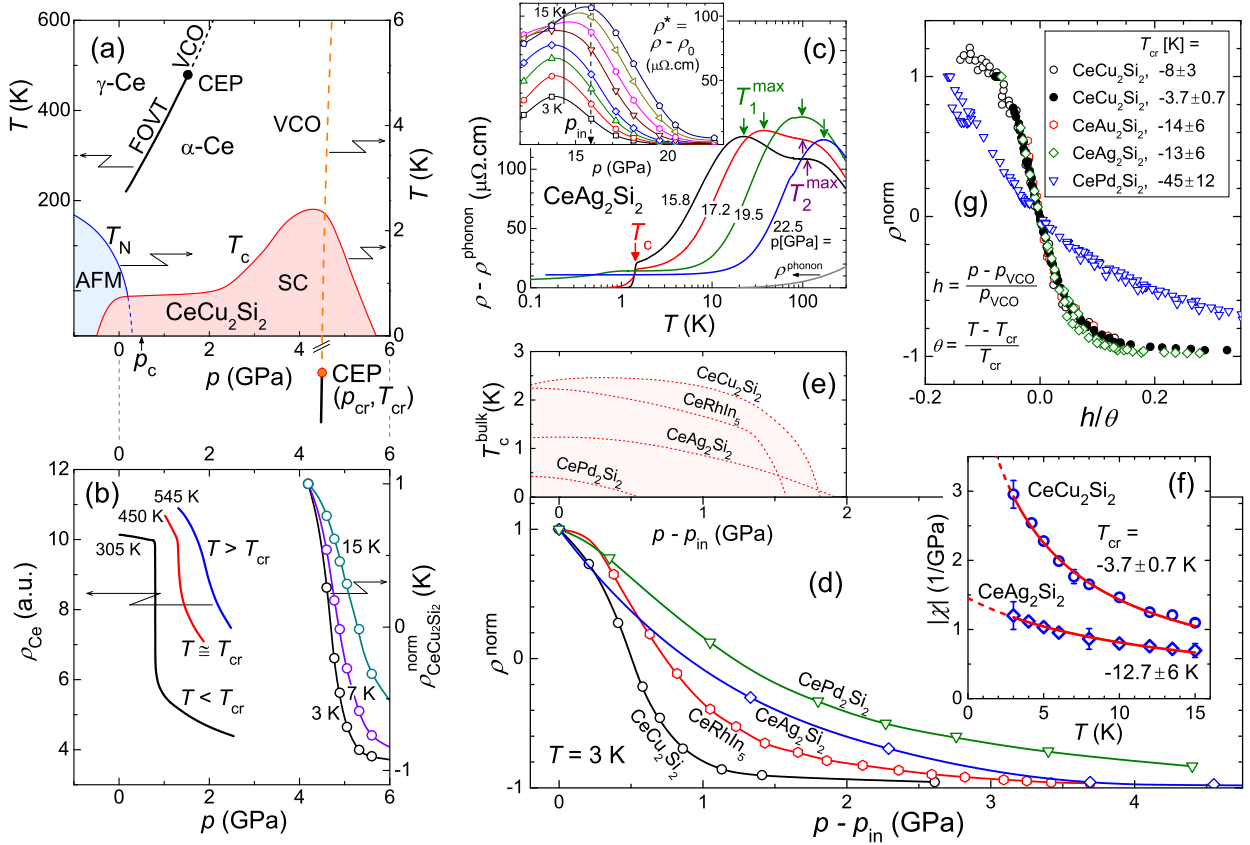

FIG. 3. **Extracting the negative temperature $T_{cr}$ of the valence transition critical end point (CEP) from low-$T$ resistivity.** (a) Schematic $p$-$T$ phase diagrams of elementary Ce [32] and CeCu$_2$Si$_2$ [8]. (b) Isothermal resistivity $\rho$ versus pressure $p$ of Ce in vicinity of the valence transition CEP [33] and normalized resistivity $\rho^{norm}$ vs $p$ of CeCu$_2$Si$_2$ in the VCO regime. (c) Resistivity $\rho - \rho^{phonon}$ versus temperature $T$ of CeAg$_2$Si$_2$ at selected pressures. Inset: Resistivity isotherms $\rho^* = \rho - \rho_0$ of CeAg$_2$Si$_2$ versus pressure at temperatures from 3 to 15 K. (d) $\rho^{norm}$ versus $p - p_{in}$ of CeCu$_2$Si$_2$, CeRhIn$_5$, CeAg$_2$Si$_2$, and CePd$_2$Si$_2$ at 3 K. The lines are guides to the eyes. (e) Schematic diagram of bulk SC for the same compounds (same $p$-scale as in (d)). (f) Slope $\chi = |\mathrm{d}\rho^{norm}/\mathrm{d}p|_{p_{VCO}}$ versus temperature $T$ of CeCu$_2$Si$_2$ and CeAg$_2$Si$_2$. The red lines represent fits to the data with $\chi \propto (T - T_{cr})^{-1}$. Error bars on $\chi$, shown for representative data points, correspond to the over- and underestimation of $\chi$ due to a low $p$-run density. (g) Normalized resistivity $\rho^{norm}$ versus the generalized distance $h/\theta$ from the CEP of CeCu$_2$Si$_2$, CeAu$_2$Si$_2$, CeAg$_2$Si$_2$, and CePd$_2$Si$_2$. (c-g) See Supplementary Table 1 for references.

For a detailed analysis, the $p$-dependence of $\rho^* = \rho - \rho_0$ at several temperatures is derived from $\rho(T)$ of successive pressure runs, as shown for CeAg$_2$Si$_2$ in Fig. 3(c). A preliminary remark is that in all Ce-HF compounds the resistivity $\rho^*$ is strongly reduced by $1 - 2$ orders of magnitude, when the system is tuned by pressure through the VCO. Such a reduction, which exceeds that expected for a progressive increase of the c-f hybridization, is attributed to a more or less sudden delocalization of 4f electrons [12]. In order to disentangle the intrinsic effect of electron delocalization from that of the temperature-dependent scattering rate, the resistivity has to be normalized. For this purpose we define an initial pressure $p_{in}$, which signals the onset of the VCO resistivity collapse (see inset of Fig. 3(c) and page 9 of the Supplementary Material for CePd$_2$Si$_2$). Then, the normalized resistivity $\rho^{norm} = \frac{\rho^*(p) - \rho^*(p_{VCO})}{\rho^*(p_{VCO})}$, where $p_{VCO}$ is the pressure of the mid drop of $\rho^*$, is calculated for each temperature.

By way of example, Fig. 3(d) shows $\rho^{norm}$ versus $p - p_{in}$ at 3 K of CeCu$_2$Si$_2$, CeRhIn$_5$, CeAg$_2$Si$_2$, and CePd$_2$Si$_2$. Clearly, it appears that the collapse of $\rho^{norm}$ is always close to optimal SC and a steeper collapse favors higher $T_c$ [see Fig. 3(e)]. With increasing temperature, i.e., increasing distance from the CEP, the pressure scale of the resistivity reduction gets broader and broader and the steepness of the collapse decreases, as shown for CeCu$_2$Si$_2$ in Fig. 3(b). Fig. 3(f) displays the temperature dependence of the slope $\chi = |d\rho^{norm}/dp|_{p_{VCO}}$ of CeCu$_2$Si$_2$ and CeAg$_2$Si$_2$. $\chi$, which we interpret as valence susceptibility, tends to diverge as $\chi \propto (T - T_{cr})^{-1}$, i.e., a first-order discontinuity would occur in $\rho^*(p)$ for $T < T_{cr}$. Evidently, a simple fit to $\chi(T)$ yields $T_{cr}$. The empirical law $(T - T_{cr})^{-1}$ is confirmed by data from several samples of CeCu$_2$Si$_2$ [8], CeAu$_2$Si$_2$ [18, 26, 27], and CeRhIn$_5$ [17], which are by the way the compounds with highest $T_c$ and least negative $T_{cr}$. Note that the plot of $\chi(T)$ is limited to a temperature (15 K), which corresponds to a small fraction of the first crystal-field-splitting energy. Such a treatment is repeated on all appropriate data from Ce-based HF compounds (see Supplementary Figs. S4–S13). The extracted $T_{cr}$ values and other parameters ($T_c$, $\rho_0$) are summarized in Table S1 of the Supplementary Material.

After identifying $T_{cr}$, one can apply the scaling treatment developed in Ref. [8] for CeCu$_2$Si$_2$ within the framework of universal scaling theory of critical phenomena and subsequently applied on data from CeAu$_2$Si$_2$ [18, 26, 27] and CeRhIn$_5$ [17]. To this end, a generalized distance $h/\theta$ from the CEP is calculated, where $h = (p - p_{VCO})/p_{VCO}$ and $\theta = (T - T_{cr})/|T_{cr}|$. Then, for a given compound, all $\rho^{norm}$ isotherms in the VCO regime collapse on a single curve $\rho^{norm} = f(h/\theta)$ when plotted versus $h/\theta$, as shown in Fig. 3(g). This means that for the generalized distance $h/\theta$ from the CEP, the $\rho^{norm}$ isotherms behave in a unique manner, which strongly supports the existence of the valence CEP at ($p_{cr}$, $T_{cr}$). Note that in terms of universal scaling theory of critical phenomena the equation is $\rho^{norm}/h^{1/\delta} = f(h/\theta^{\gamma/(\delta-1)})$, with the critical exponents $\gamma$ and $\delta$ (mean-field approach:

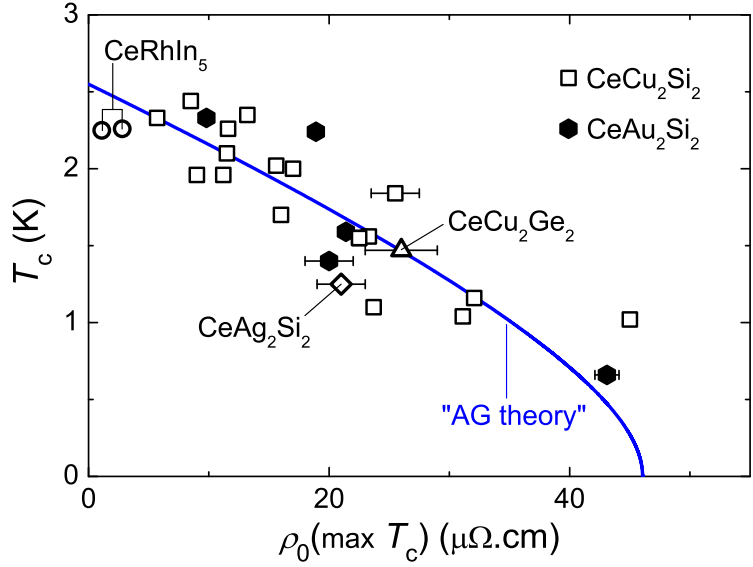

FIG. 4. **Pair-breaking systematics:** $T_c$ vs $\rho_0$ ($\rho_0$ taken at the pressure of maximal $T_c$) of the CeCu$_2$Si$_2$ family and CeRhIn$_5$ (published and unpublished data, see Supplementary Fig. S1 for references). The error bars on $\rho_0$ result from the power-law extrapolation to zero temperature of $\rho(T)$. The solid line represents the generalized Abrikosov-Gor'kov theory with a critical resistivity $\rho_0^{cr} = 46$ $\mu\Omega$cm [38].

$\gamma_{MF} = 1$, $\delta_{MF} = 3$). Accordingly to Ref. [8], the critical exponents are fixed as $\gamma = 1$ and $\delta \to \infty$, which does not correspond to a simple universality class.

The scaling is very robust for 15 data sets from 7 different systems and astonishingly the scaling function $f(h/\theta)$ is identical for the isovalent systems CeCu$_2$Si$_2$, CeAg$_2$Si$_2$, and CeCu$_2$Si$_2$ [see Fig. 3(g)]. However, $f(h/\theta)$ is different for CePd$_2$Si$_2$ and apparently material dependent. For instance $f(h/\theta)$ of CeCu$_5$Au (Supplementary Fig. S12) lies in between the two data sets of Fig. 3(g). The scaling fails for two data sets: the $\rho^{norm}$ isotherms of CeCu$_6$ (Supplementary Fig. S11) and CeIrIn$_5$ (Supplementary Fig. S13) do not collapse on a single curve. Measurement errors surely play a role for the CeCu$_6$ sample (see Supplementary Fig. S11) and a change of regime ascribed to the crystal field effect [34] may interfere. In CeIrIn$_5$, low-temperature resistivity properties hint to a pressure-induced change of regime (see Supplementary Fig. S13), which may explain that the isotherms collapse only for $h/\theta > 0$. Though, completely satisfactory explanations are still missing for both.

We now discuss the pair-breaking effect by disorder, which is quantified by the residual resistivity $\rho_0$. Figure 4 shows $T_c$ versus $\rho_0$ of the CeCu$_2$Si$_2$ family and CeRhIn$_5$ from all 26 independent pressure experiments done in Geneva with $\rho_0 < 50$ $\mu\Omega$cm (see Supplementary

Fig. S1 for details and references). Not included are still finite $T_c$ values corresponding to very high $\rho_0$, which deviate from the general trend possibly due to alloying or Kondo-hole effects. For instance, in a $CeCu_2(Si_{1-x}Ge_x)_2$ alloy, a maximum bulk $T_c \sim 0.6$ K ($\rho = 0$ criteria) is reported for $\rho_0 \sim 70$ $\mu\Omega$cm [35], and, in polycrystalline $CeCu_2Si_2$, a maximum bulk $T_c \sim 0.3$ K is given with $\rho_0 \sim 180$ $\mu\Omega$cm, which is higher than the room temperature resistivity [36]. The evident decrease of $T_c$ with increasing disorder follows qualitatively well the formula given by the Abrikosov-Gor'kov (AG) theory [37] generalized for non-magnetic disorder in a CVF-mediated d-wave superconductor with a critical resistivity $\rho_0^{cr} = 46$ $\mu\Omega$cm [38]. Note that every data point in Fig. 4 refers to a set of pressure runs for an experiment on a sample of specific quality as reflected by its $\rho_0(p = 0)$ value. Thus in spite of a given data scattering, the systematic dependence of $T_c$ on $\rho_0$ for different samples of different compounds is remarkable.

A similar trend is observed for the $CePd_2(Si/Ge)_2$ family, where bulk SC vanishes completely for $\rho_0$ higher than 3 $\mu\Omega$cm (see Supplementary Fig. S2). $T_c$ is already small at ideal sample quality since these compounds are located far from the criticality as signaled by the large negative $T_{cr} \sim -50$ K. Therefore, the theory of Okada $et$ $al.$ [38] for robustness of $T_c$ versus pair-breaking effect by non-magnetic disorder is not appropriate, and the conventional AG-theory for anisotropic superconductivity can be applied. The latter is valid for the d-wave order parameter predicted by the CVF theory [23] and accounts for the rapid decrease of $T_c$ in $CePd_2Si_2$.

The critical resistivity $\rho_0^{cr} \approx 46$ $\mu\Omega$cm of the "high-$T_c$" HF superconductors is far larger than that expected in the conventional case of weak-coupling SC but is compatible with the generalized AG theory [38]. In fact, due to the valence fluctuation renormalization effect of the impurity potential [39], $\rho_0$ is strongly increased at pressures around $p_{cr}$ compared to far lower or higher pressures, which is a hallmark of CVF-mediated HF superconductors (exceptions are $CeCu_6$ [34] and $CePd_2Si_2$ [19]). The robustness of $T_c$ against impurity scattering is due to the fact that the re-normalized impurity potential is a long-range like bare Coulomb potential [38, 39]. Indeed, almost all scattering channels with angular momentum $\ell = 0, 1, 2, ...$, i.e., s-, p-, d-wave and so on, are active in the Coulomb-type potential, leading to partial cancellation in the scattering rate among the $\ell$-wave vertex part in the pair susceptibility and the self-energy part in the Green function. This rationalizes the robustness of $T_c$ against the enhanced impurity potential in contrast to the conventional

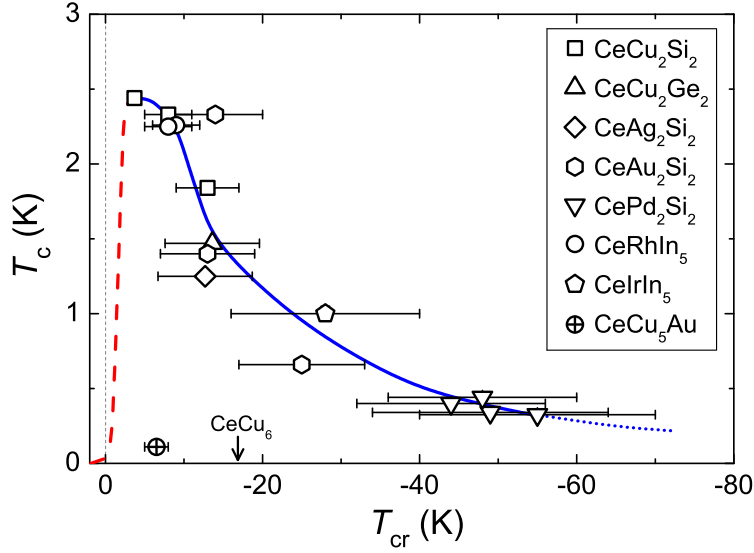

FIG. 5. **Building superconductivity up in heavy fermion compounds.** The presented $T_c$-vs-$T_{cr}$ relation is qualitatively predicted by CVF theory [23, 24]. The continuous line is a guide to the eyes. The dashed line represents a qualitative prediction from CVF theory [23]. The error bars on $T_{cr}$ represent estimated errors according the scaling analysis (see main text). The arrow indicates $T_{cr}$ of CeCu$_6$ (no SC).

AG-type theory for anisotropic pairing, where essentially an s-wave component of impurity potential is taken into account [40, 41].

Figure 5 presents the most interesting relationship between $T_c$ and the key parameter $T_{cr}$. $T_c$ is maximal for small negative $T_{cr}$ and decreases as $T_{cr}$ becomes more negative, which is qualitatively predicted by CVF theory [23, 24]. About half of the data correspond to samples with almost optimal $T_c$ for the specific compounds thanks to low $\rho_0$ values (see Supplementary Table S1), which underlines the intrinsic character of the $T_c$-vs-$T_{cr}$ relationship. Unlike Fig. 4, less data points are presented because of the stringent requirement of accuracy and reliability of resistivity measurement for the scaling analysis, which excludes a part of our results and also those found in literature. In respect of the procedure to deduce $T_{cr}$, a prerequisite is the accurate control of the absolute resistivity value as function of pressure (form factor) and temperature and a high pressure run density. The control of the form factor is far to be an easy task in high-pressure cells and main complication comes from non-hydrostatic conditions in various pressure transmitting medium such as He [42, 43], Daphne oil [8, 25], or steatite [18, 44]. Another requirement is the limitation of non-systematic error on pressure and the precise estimation of $\rho_0$. Moreover, due to the presumed $1/(T - T_{cr})$ dependence of $\chi$, the uncertainty on $T_{cr}$ is magnified for large negative values, while the smallest values are the most accurate. Namely, in the case of $CeCu_2Si_2$ with $T_{cr} = -3.7$ K, the error is within the symbol size.

Despite reliable small $T_{cr}$ and moderate $\rho_0$, $CeCu_6$ (no SC) and $CeCu_5Au$ (partial SC at 0.11 K) clearly lie below the general trend. From a literature review, small $T_{cr}$ values should also be expected for some compounds including $CeAl_2$ [45], $CeAl_3$ [46, 47], and $CeInCu_2$ [48]. However, SC has never been observed in these cases. We have no satisfactory explanation yet for this discrepancy.

The dashed red curve for $T_{cr} \to 0$ in Fig. 5 is drawn from the theoretical prediction of the 3D model [23] in which $T_c$ is paradoxically suppressed just at the critical point of the valence transition, while $T_c$ takes sharp maximum near the VCO line in the Kondo regime. This aspect has also been verified by the density-matrix-renormalization-group calculation for the 1D model [24], which is numerically accurate. Namely, inter-site pairing correlation dominates over spin density wave and charge density wave correlations near the sharp VCO inside the Kondo regime.

## III. DISCUSSION

The systematic behavior of $T_c$ versus $\rho_0$ and $T_{cr}$ points to a possible maximum $T_c \approx 2.5$ K in Ce-based HF superconductors and strongly supports that CVF provide the dominant pairing mechanism. Although, the relation between $T_c$ and $T_{cr}$ was already inferred in the pioneer work of Onishi and Miyake [23], a quantitative prediction seems almost impossible at the present state of art. Theory also considers that $T_{cr}$ is influenced by disorder in general, which is less evident in the experimental data (see Supplementary Fig. S3). Naively, one can imagine that disorder induces an additional smearing of the VCO, which depresses $T_{cr}$ and then $T_c$.

Let us now comment some generalities about magnetism and superconductivity in HF compounds. A hallmark is the merging close to $p_{cr}$ of the two maxima in $\rho(T)$ [see Fig. 3(c)], which indicates that the rapidly rising Kondo energy starts to exceed the first crystal-field-splitting energy [49]. For HF superconductors, this pressure corresponds to optimal SC without exception. Crossing the VCO, the ground state degeneracy of the Ce ion increases from $n = 2$ to full degeneracy $n = 6$ of the 4f multiplet [7], and at the pressure of maximal $T_c$, the energy scale $T_K$ is much larger than the magnetic ordering temperature. Moreover, the strength of the f-c hybridization seems to control the position of $p_c$ in respect to $p_{cr}$ as shown theoretically [10]. In the case of strong hybridization $p_c$ and $p_{cr}$ are well separated, but in the case of weak hybridization a hypothetical magnetic QCP would occur at pressure higher than $p_{cr}$ and the VCO drives a first-order collapse of magnetism at $p_c \sim p_{cr}$ [10, 50] in parallel to the traditional competition between the RKKY and Kondo energies. Hence the collapse of magnetism is very abrupt or even first-order-like in CeAu$_2$Si$_2$ [18, 51], CeAg$_2$Si$_2$ [52], and CeRhIn$_5$ [53].

Although pressure is a clean tuning parameter, clear evidence of a second-order magnetic transition down to zero temperature and a resulting quantum critical point is not well established from pressure investigations on pure lattices. In the particular case of CePd$_2$Si$_2$, results of Refs. [3, 43, 54, 55] support a linear decrease of $T_N$ down to zero with $p$ approaching $p_c$. However, a more rapid vanishing of $T_N$ appears to correspond to higher superconducting $T_c$ [19, 56] (see Supplementary Fig. S14). The difficulty of tracking the $T_N$ vanishing with resistivity or even heat capacity probes and the unavoidable pressure gradient, due to which the $T_N$ decrease appears more progressive, impedes a clear-cut conclusion. Seemingly second-order-like magnetic collapses have only been established for alloy systems, see e.g. [35, 57] and a lattice [58] with relatively high $\rho_0$ value, where disorder likely masks the intrinsic behavior.

The overlap of magnetic order and SC in CeRhIn$_5$ [50, 53] and especially in CeAu$_2$Si$_2$ [18] (see Fig. 1) contradicts the longstanding consensus that HF SC emerges in the vicinity of the magnetic border [4]. From a global point of view, the CeCu$_2$Si$_2$ family shows quite different magnetic phase diagrams concomitantly with otherwise strikingly similar electric and thermoelectric transport and superconducting properties [18, 27, 51]. For instance, a systematic feature in thermopower precedes the occurrence of SC [27]. Consequently, the link between SC and magnetism is primarily a competition, with the possible exception of CeAu$_2$Si$_2$ [18]. Up to a given delocalization of 4f electrons, magnetism hinders CVF to build up SC. On the other hand, the low-pressure SC pocket in CeCu$_2$Si$_2$ is the best candidate for spin fluctuation mediated SC [59], because the magnetic collapse at $p_c \approx 0$ and the VCO at $p_{cr} \approx 4.2$ GPa are exceptionally well separated. Though, the scenario of single-band nodal-d-wave SC at $p = 0$ in CeCu$_2$Si$_2$ is now strongly challenged [60–63].

Let us comment on the Kondo-volume-collapse mechanism introduced by Razafimandimby $et$ $al.$ [64]. To our understanding, it is essentially a phonon-mediated SC mechanism due to enhanced electron-phonon coupling through the Kondo-volume-collapse effect (large Gruneisen parameter). In this regard, it should be different from the valence fluctuation mediated mechanism. According to an almost exact (justified by the Ward identity argument) theoretical discussion based on periodic Anderson model with coupling to phonon by Jichu $et$ $al.$ [65], it seems rather difficult for this mechanism to build up "high-$T_c$" SC in Ce-based heavy fermions. According to Jichu $et$ $al.$, the enhanced pairing interaction (by Kondo-volume-collapse effect) vanishes at the static limit. Furthermore, it is crucial to note that the valence fluctuation mechanism is not based on density fluctuations but fluctuations of f-c charge transfer with the total charge density ($n_f + n_c$) essentially kept constant. Namely, valence fluctuations are rather categorized with orbital fluctuations.

Finally, we comment on CVF in Yb-based HF compounds, which can be approached as electron-hole/inverse-pressure analogues of Ce compounds. Interestingly, the first discovered Yb-based superconductor $\beta$-YbAlB$_4$ [66] exhibits normal state properties with unconventional quantum criticality [67], which is naturally explained by the CVF theory [68]. Furthermore, common criticality has been observed in some classes of Yb-based periodic crystals and even in the quasicrystal Yb$_{15}$Au$_{51}$Al$_{34}$ [69]. Search for superconductivity induced by CVF in Yb-based systems as well as the identification of the CEP of the underlying Yb-valence transition on the basis of the method described in this paper is expected to open a new frontier in this field.

For a long time, the spin-fluctuation-mediated mechanism was the mainstream scenario for SC in HF systems. However, the CVF theory has provided a new framework able to account for the high-pressure superconducting phase and several other phenomena in CeCu$_2$Si$_2$. We now have shown that this theory is able to explain salient experimental features in a multitude of systems, corroborating CVF as a plausible Cooper pairing mechanism. Concretely, the present study provides striking evidence that the optimum superconducting $T_c$ in many Ce-based HF superconductors is essentially controlled by the strength of CVF and by non-magnetic disorder. Furthermore, we believe that CVF-induced SC is connected to a much wider part of non-trivial physics in strongly correlated electron systems including high-$T_c$ cuprates [21, 70, 71]. Thus, our work uncovers a new playground for condensed matter physicists.

**METHODS**

The above presented results are based on electrical resistivity data obtained on 17 single crystals from 9 different Ce-based HF compounds (see Supplementary Table S1). The four-point electrical resistivity measurements under high-pressure conditions have been carried out in standard helium and dilution cryostats. The high-pressure conditions where obtained using Bridgman pressure cells with tungsten-carbide or diamond anvils and with different pressure transmitting medium. Technical details can be found for each transmitting medium in (helium) [7, 42], (Daphne oil) [8, 25], and (steatite) [18, 44]. All relevant information about crystals growth, dimension of sample and pressure cell, and data acquisition can be found in the respective references (see Supplementary Table S1). All relevant information about data treatment can be found in Ref. [8] and in the main text.

**SUPPLEMENTARY MATERIAL**

The Supplementary Material follows after the bibliography:
- Table S1
- Figures S1 to S14

**ACKNOWLEDGMENTS**

We acknowledge H. Wilhelm, S. Raymond, and A. T. Holmes for enabling the analysis of previous data, and M. Lopez, S. Müller, and M. Tran for technical support. D. Aoki acknowledges for the financial support KAKENHI (15H05882, 15H05884, 15K21732, 16H04006). K. Miyake acknowledges for the financial support KAKENHI (17K05555). S. Watanabe acknowledges for the financial support KAKENHI (15K05177, 16H01077).

**AUTHOR CONTRIBUTIONS**

D. J. conceived the idea and designed the experiment. D. A. and G. L. have grown the single crystal samples. G. S. and Z. R. performed the measurements. G. S. analyzed the data. S. W. and K. M. provided the microscopic-theoretical interpretation. G. S. and D. J. wrote the paper. All authors discussed the results and commented on the manuscript.

# Supplementary Material for

## The Dominant Role of Critical Valence Fluctuations on High $T_c$ Superconductivity in Heavy Fermions

Gernot W. Scheerer*, Zhi Ren, Shinji Watanabe, Gérard Lapertot, Dai Aoki, Didier Jaccard, Kazumasa Miyake

*correspondence to: gernot.scheerer@unige.ch

**This file includes:**

| compound | $\rho_0(p = 0)$ [μΩcm] | $\rho_0$ (maximum $T_c$) [μΩcm] | $T_{cr}$ [K] | maximum $T_c$ [K] | $\rho$-data first published | $T_{cr}$ from |
|---|---|---|---|---|---|---|
| $CeCu_2Si_2$ | ~ 5 | 25.5(2.0) | -13(4) | 1.84 | [42] | Fig. S4 |
| $CeCu_2Si_2$ | 0.4 | 5.7(0.2) | -8(3) | 2.33 | [8] | [8] |
| $CeCu_2Si_2$ | see Fig. S5 | 8.5(0.2) | -3.7(7) | 2.44 | Fig. S5 | Fig. S5 |
| $CeCu_2Ge_2$ | ~ 2 | 26(3) | 13.6(6.0) | 1.47 | [72] | Fig. S6 |
| $CeAg_2Si_2$ | 5.5 | 21(2) | 12.7(6.0) | 1.25 | [52] | [52] |
| $CeAu_2Si_2$ | 1.8 | 9.8(0.2) | -14(6) | 2.33 | [18] | [18] |
| $CeAu_2Si_2$ | 2.5 | 20(2) | -13(6) | 1.40 | [27] | [27] |
| $CeAu_2Si_2$ | 12 | 43(1) | -25(8) | 0.66 | [26] | [26] |
| $CePd_2Si_2$ | ~ 4 | 1.93(0.05) | -44(12) | 0.40* | [56] | Fig. S7 |
| $CePd_2Si_2$ | 2.6 | 1.83(0.05) | -49(15) | 0.34* | [56] | Fig. S8 |
| $CePd_2Si_2$ | 1.25 | 1.1(0.05) | -48(12) | 0.441 | [19] | Fig. S9 |
| $CePd_2Si_2$ | 0.6 | 0.44(0.05) | -55(15) | 0.325 | [19] | Fig. S10 |
| $CeCu_6$ | 12.4 | --† | -17(9) | --† | [34] | Fig. S11 |
| $CeCu_5Au$ | 28.9 | 33(0.5) | -6.5(1.5) | 0.11** | [75] | Fig. S12 |
| $CeRhIn_5$ | $0.2^{T = 2\,K}$ | 2.8(0.1) | -9(3) | 2.26 | [17] | [17] |
| $CeRhIn_5$ | 0.011 | 1.1(0.05) | -8(3) | 2.25 | [17] | [17] |
| $CeIrIn_5$ | $1.7^{T = 1.5\,K}$ | 0.37(0.02) | -28(12) | 1 | Fig. S13 | Fig. S13 |

**Table S1**. Experimental values of $\rho_0(p = 0)$, $\rho_0$ taken at the pressure of maximum $T_c$, $T_{cr}$, and maximum $T_c$ of several Ce-HF samples. Note that "maximum $T_c$" refers to the maximum value of the bulk-superconducting transition temperature as function of pressure in a given sample, except for two $CePd_2Si_2$* samples and $CeCu_5Au$**. In general, bulk superconductivity coincides with zero resistivity, in agreement with ac heat capacity or magnetic susceptibility signals [8,18,51,73]. Resistivity $\rho$ was measured in the basal plane or along the a-axis of the tetragonal structure, except for $CeCu_6$ and $CeCu_5Au$, where $\rho$ was investigated for the b-axis of the orthorhombic structure. *$T_c$ from resistivity transition offset criterion, where $\rho$ does not completely drop to zero [56]. **$T_c$ from resistivity transition onset criterion [75]. † By matching the ($p$-$T$) phase diagrams of $CeCu_6$ and $CeCu_5Au$, hypothetical SC in $CeCu_6$ corresponds to negative pressure $p \approx$ -0.5 GPa.

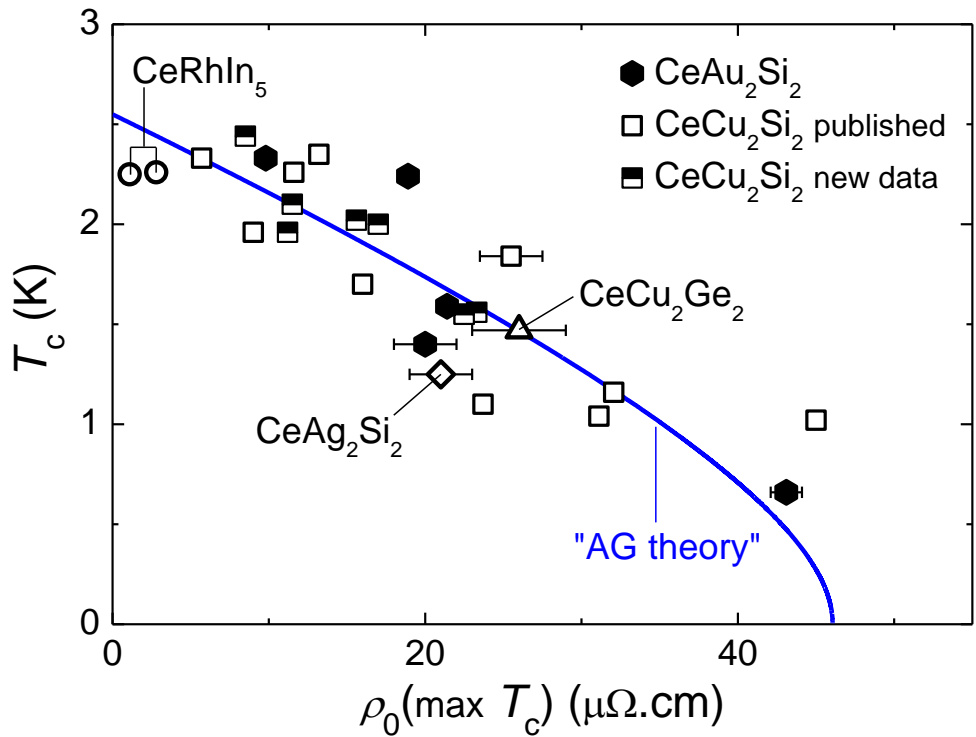

**Fig. S1.** Superconducting $T_c$ vs. residual resistivity $\rho_0$ of the $CeCu_2Si_2$-family and $CeRhIn_5$ ($\rho_0$ taken at the pressure of maximum $T_c$). All data points are from independent pressure experiments done in Geneva. Data for $CeRhIn_5$ from [17], $CeCu_2Ge_2$ from [72], and $CeAg_2Si_2$ from [52]. Filled hexagons refer to published data of $CeAu_2Si_2$ [18,26,27,51]. Empty cubes refer to published data of $CeCu_2Si_2$ [7,8,36,42,44,73,74]. Half-filled cubes refer to new data of $CeCu_2Si_2$.

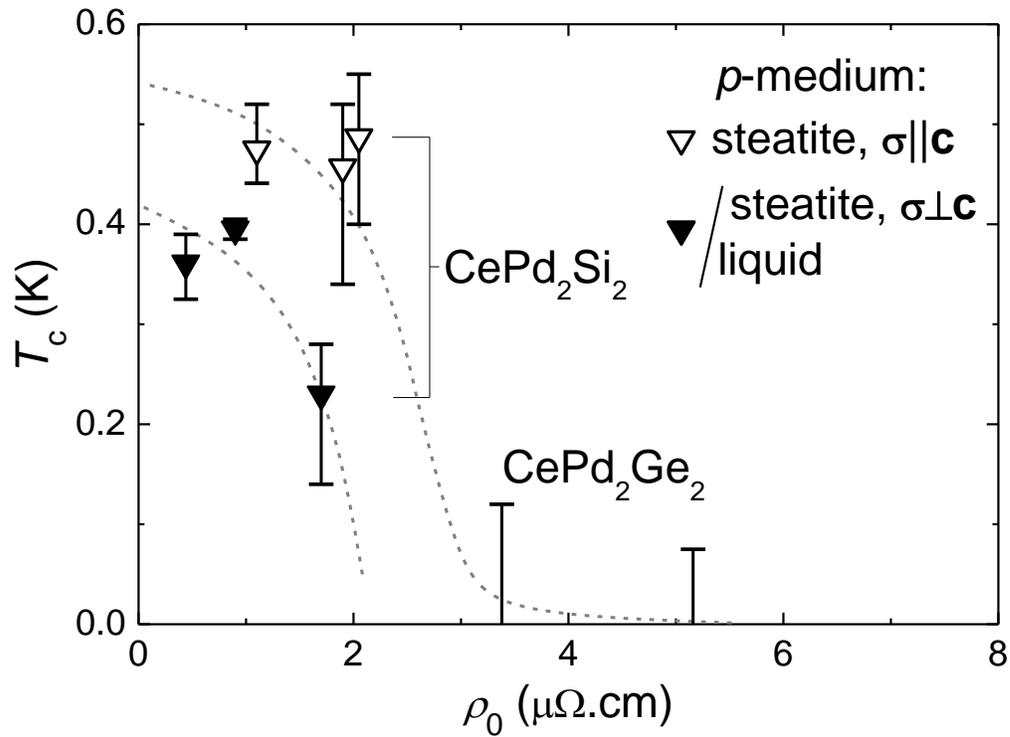

**Fig. S2.** Superconducting $T_c$ vs. residual resistivity $\rho_0$ of CePd$_2$Si$_2$ [19,43,55,56] and CePd$_2$Ge$_2$ [76] ($\rho_0$ taken at the pressure of maximal $T_c$). Triangle symbols indicate the $T_c$ defined at the 50% drop of $\rho$. The onset and end of the resistive superconducting transition are indicated by vertical bars. The dashed lines are guides to the eyes. An enhanced $T_c$ due to strain effects [19] is observed in soft-solid steatite pressure medium with $\boldsymbol{\sigma} \parallel \mathbf{c}$ compared to liquid pressure medium or steatite with $\boldsymbol{\sigma} \perp \mathbf{c}$, and two data sets with qualitatively similar behavior can be distinguished.

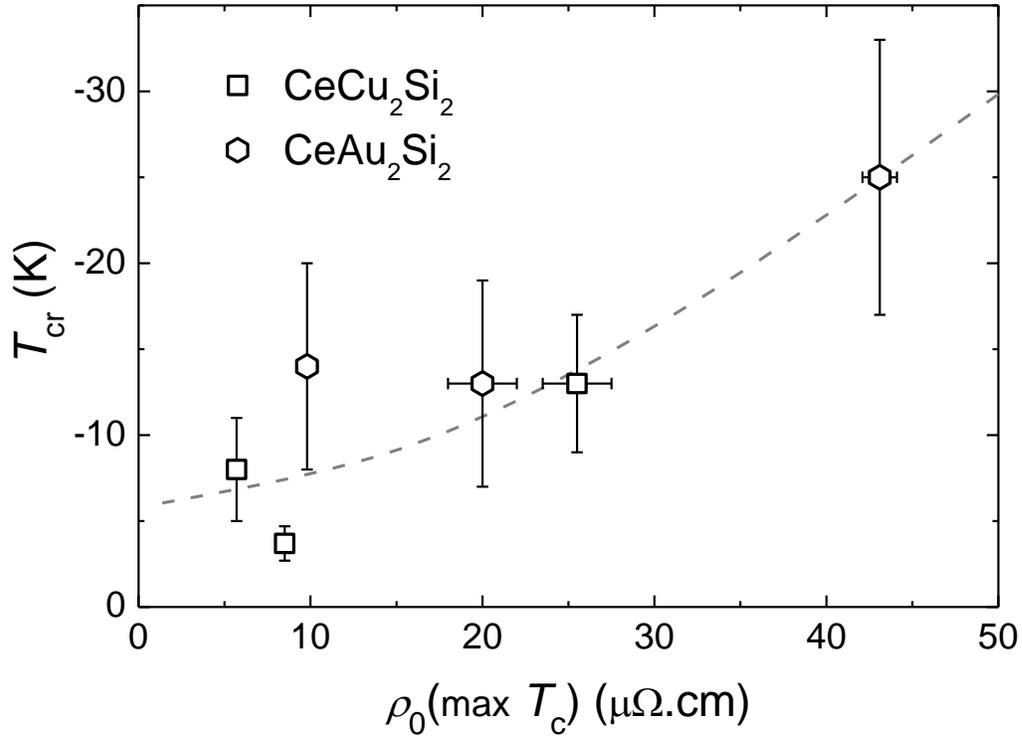

**Fig. S3.** Critical temperature $T_{cr}$ vs. residual resistivity $\rho_0$ of $CeCu_2Si_2$ and $CeAu_2Si_2$ ($\rho_0$ taken at the pressure of maximal $T_c$). As qualitatively predicted by CVF theory [23], the two parameters $\rho_0$ and $T_{cr}$ are correlated. Disorder induces an additional smearing of the VCO-induced resistivity collapse resulting in reduced experimental $T_{cr}$ value.

**Supplementary Figure S4**

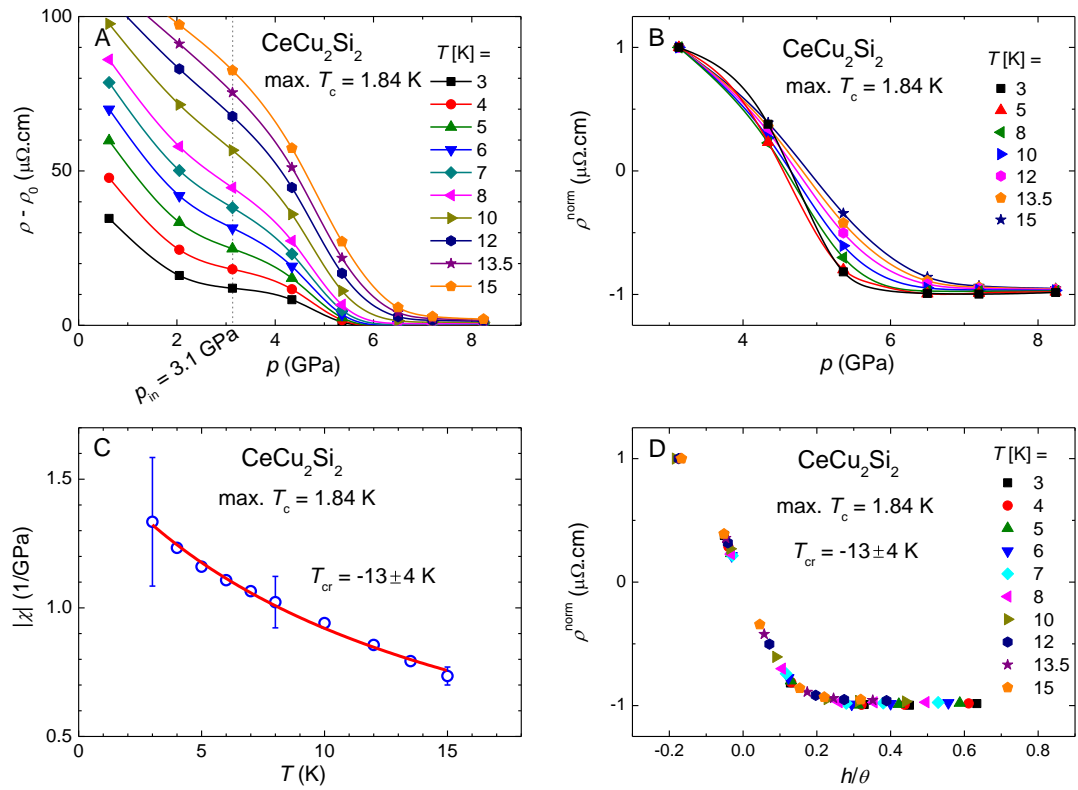

**Fig. S4.** Resistivity scaling analysis of a CeCu$_2$Si$_2$ sample with $T_c = 1.84$ K. $\rho(T)$-data first published in [42]. (A) Resistivity isotherms $\rho^* = \rho - \rho_0$ vs. pressure $p$ at temperatures $T$ from 3 up to 15 K. (B) Normalized resistivity $\rho^{\text{norm}}$ vs $p$. (C) Slope $\chi = (\mathrm{d}\rho^{\text{norm}}/\mathrm{d}p)_{p\text{VCO}}$ vs. $T$. The red line represents a fit to the data with $\chi \sim 1/(T - T_{\text{cr}})$. The fit gives $T_{\text{cr}} = -13 \pm 4$ K. Error bars on $\chi$ are shown for representative data points. (D) Normalized resistivity $\rho^{\text{norm}}$ vs. the generalized distance $h/\theta$.

**Supplementary Figure S5**

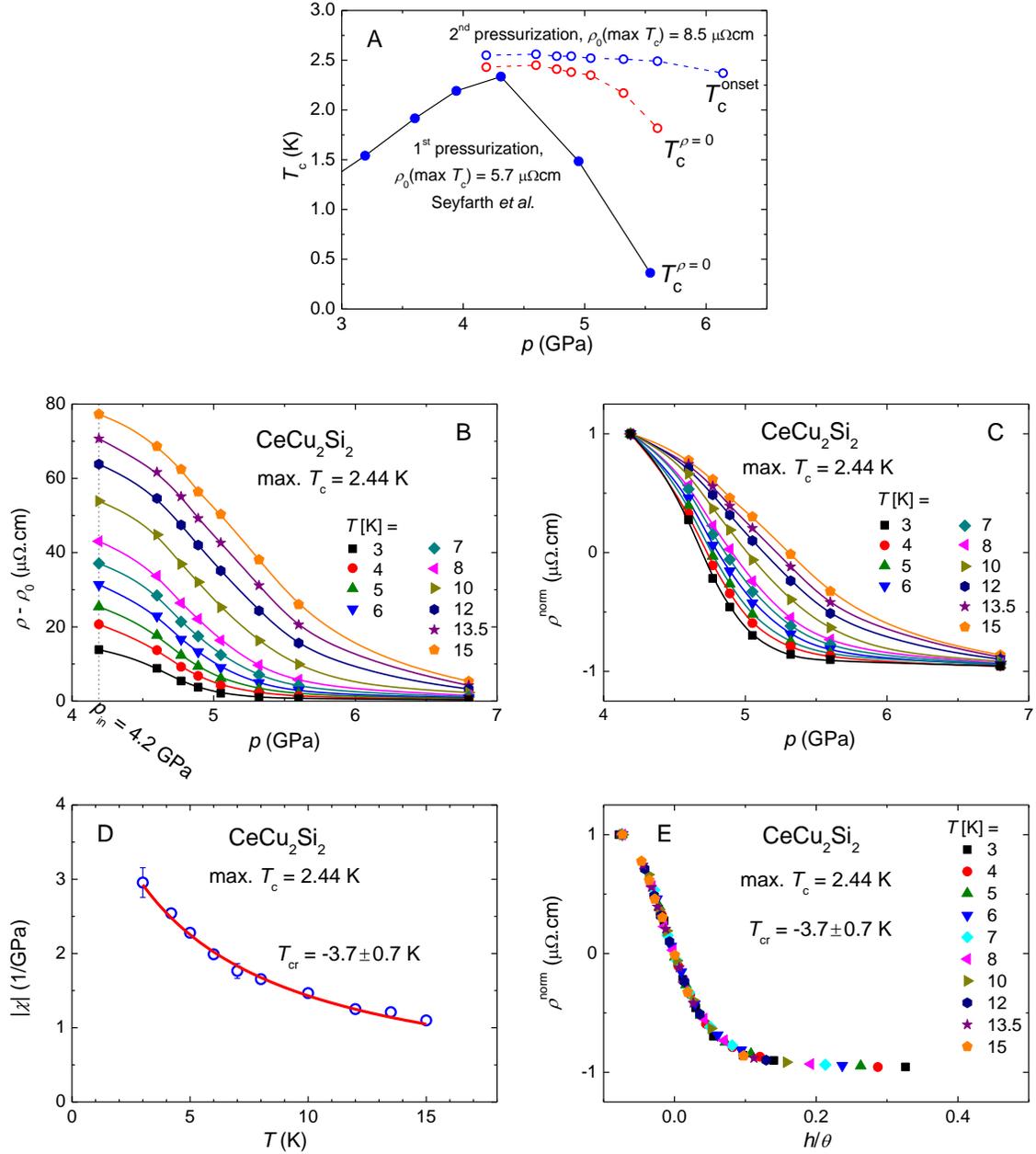

**Fig. S5.** Resistivity scaling analysis of a CeCu$_2$Si$_2$ sample with $T_c = 2.44$ K (record value!). Unpublished data of the second pressure cycle of Seyfarth *et al.* [8]. After a partial depressurization down to about 4 GPa, pressure was increased again by small steps for a better investigation of the VCO. (A) $T_c$ vs. pressure of the first and second pressure cycles. Interestingly, the maximal $T_c$ of the second cycle was a bit higher and the SC dome shifted up by about 0.5 GPa associated to a 50% increase in $\rho_0$. (B) Resistivity isotherms $\rho^* = \rho - \rho_0$ vs. pressure $p$ at temperatures $T$ from 3 up to 15 K. (C) Normalized resistivity $\rho^{norm}$ vs $p$. (D) Slope $\chi = (d\rho^{norm}/dp)_{p \text{VCO}}$ vs. $T$. The red line represents a fit to the data with $\chi \sim 1/(T - T_{cr})$. The fit gives $T_{cr} = -3.7 \pm 0.7$ K. Error bars on $\chi$ are shown for representative data points. (E) Normalized resistivity $\rho^{norm}$ vs. the generalized distance $h/\theta$. With a high pressure-run density in the VCO regime, these data give the most accurate $T_{cr}$ value.

**Supplementary Figure S6**

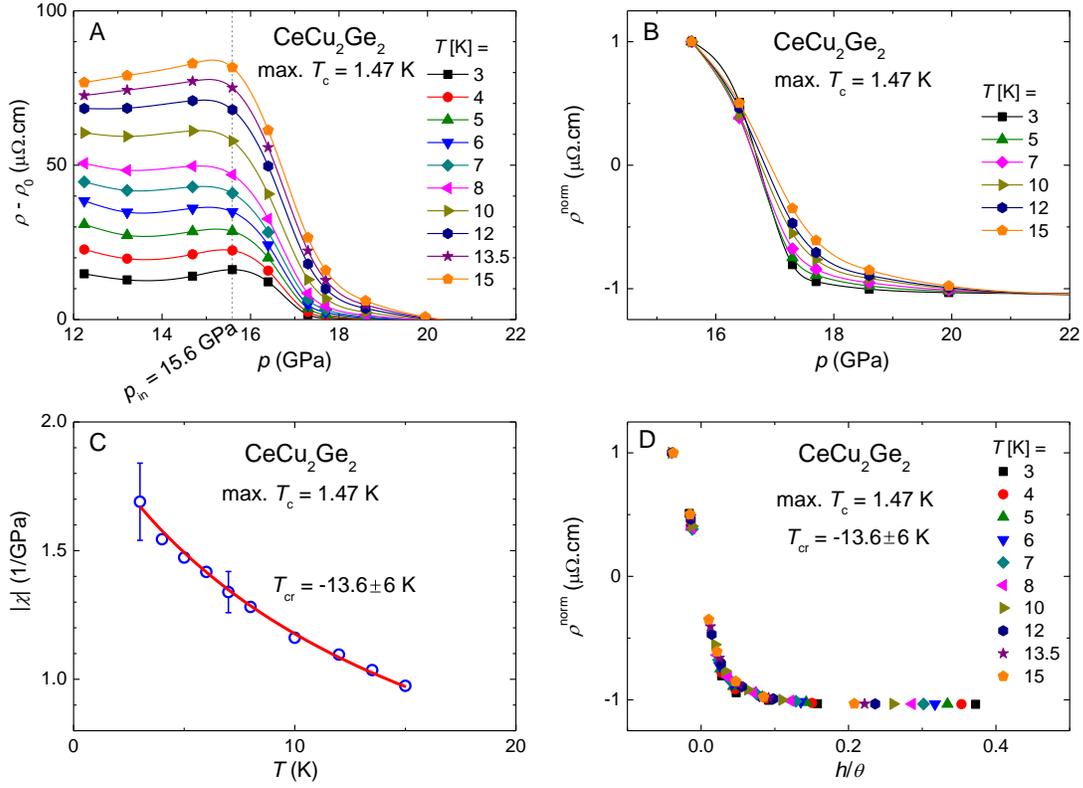

**Fig. S6.** Resistivity scaling analysis of a CeCu$_2$Ge$_2$ sample with $T_c$ = 1.47 K. $\rho(T)$-data first published in [72]. (A) Resistivity isotherms $\rho^* = \rho - \rho_0$ vs. pressure $p$ at temperatures $T$ from 3 up to 15 K. (B) Normalized resistivity $\rho^{norm}$ vs $p$. (C) Slope $\chi = (d\rho^{norm}/dp)_{pVCO}$ vs. $T$. The red line represents a fit to the data with $\chi \sim 1/(T - T_{cr})$. The fit gives $T_{cr} = -13.6 \pm 6$ K. Error bars on $\chi$ are shown for representative data points. (D) Normalized resistivity $\rho^{norm}$ vs. the generalized distance $h/\theta$.

**Comment on scaling analysis of CePd₂Si₂:**

From the "raw" resistivity isotherms $\rho^*$ (obtained in steatite $p$-medium) it is obvious that the resistivity collapse is slower than in the other compounds (see main text Fig. 2). Thus, the CEP should be at more negative temperature, which would imply a slower change of the slope $\chi$ as a function of $T$, and a bigger error on $T_{cr}$ extracted by fitting with $\chi \sim 1/(T - T_{cr})$. Fortunately in CePd₂Si₂, the relatively high value $T_1^{max} \sim 100$ K at the pressure of the VCO allows to extend the analysis to higher temperature. By fitting $\chi(T)$ over an extended $T$-scale, the error of $T_{cr}$ is reduced. Very robust scaling analysis is obtained for four CePd₂Si₂ samples (see Figs. S7-S10).

With a CEP at very negative temperature in CePd₂Si₂, the decrease of $\rho^*$ is extended over a larger pressure scale compared to e.g. CeCu₂Si₂. For $T > T_N$, the decrease of $\rho^*$ sets in already at pressures well below $p_c$. We have chosen $p_{in}$ with the condition $p_{in} > p_c$ to go as low as possible in temperature to get the closest to the CEP. We have verified that the conditions $p_{in} < p_c$ and $T > T_N$ give similar results, with $T_{cr}$ values within the error bar.

**Supplementary Figure S7**

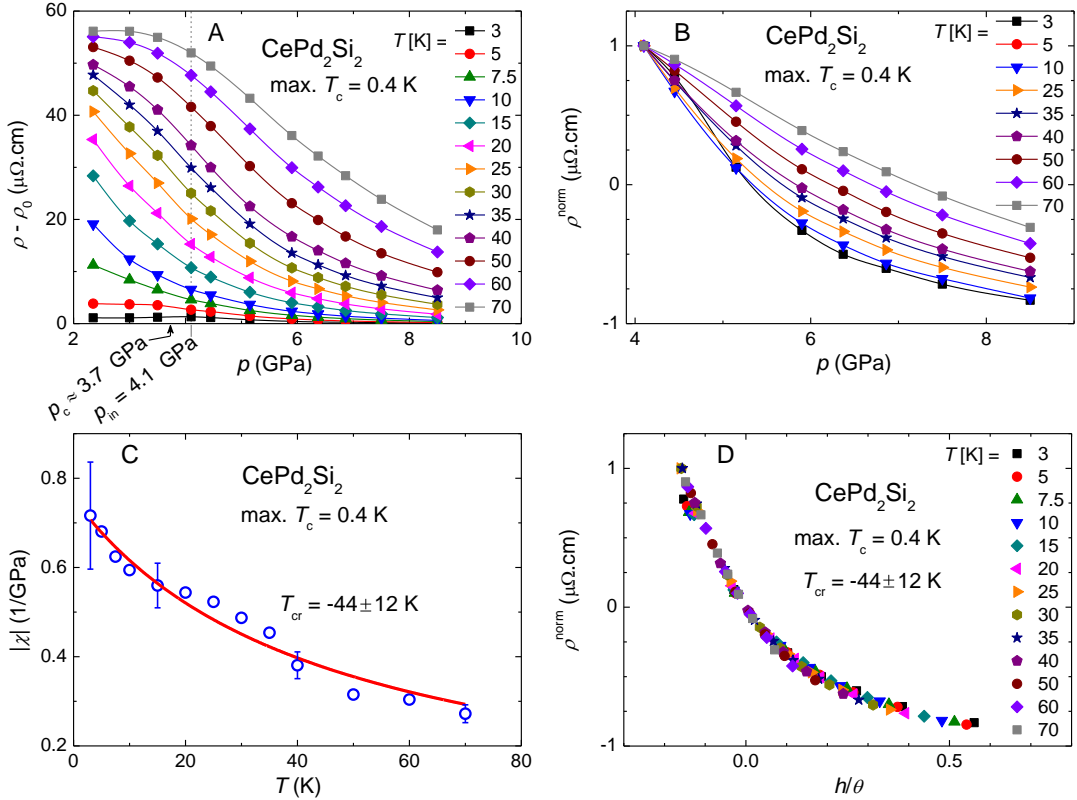

**Fig. S7.** Resistivity scaling analysis of a CePd₂Si₂ sample with $T_c = 0.4$ K. $\rho(T)$-data first published in [56]. (A) Resistivity isotherms $\rho^* = \rho - \rho_0$ vs. pressure $p$ at temperatures $T$ from 3 up to 70 K. (B) Normalized resistivity $\rho^{norm}$ vs. $p$. (C) Slope $\chi = (d\rho^{norm}/dp)_{p\text{VCO}}$ vs. $T$. The red line represents a fit to the data with $\chi \sim 1/(T - T_{cr})$. The fit gives $T_{cr} = -44 \pm 12$ K. Error bars on $\chi$ are shown for representative data points. (D) Normalized resistivity $\rho^{norm}$ vs. the generalized distance $h/\theta$.

**Supplementary Figure S8**

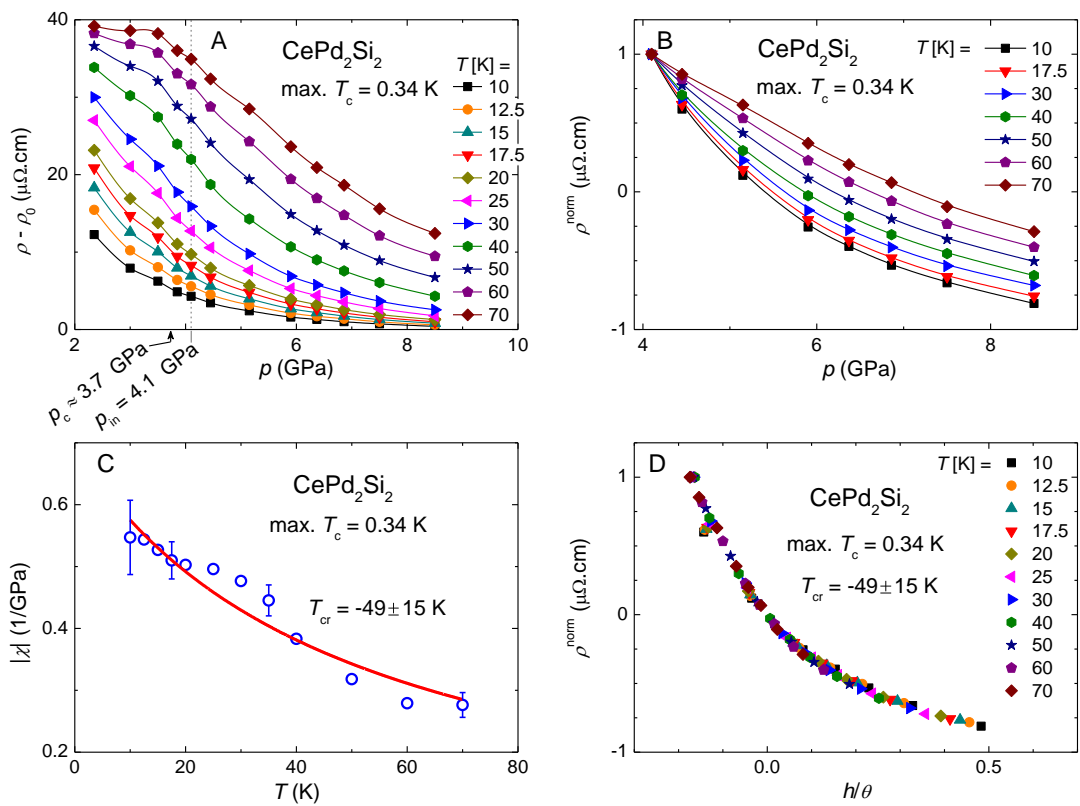

**Fig. S8.** Resistivity scaling analysis of a CePd$_2$Si$_2$ sample with $T_c$ = 0.34 K. $\rho(T)$-data first published in [56]. (A) Resistivity isotherms $\rho^* = \rho - \rho_0$ vs. pressure $p$ at temperatures $T$ from 10 up to 70 K. (B) Normalized resistivity $\rho^{norm}$ vs $p$. (C) Slope $\chi = (d\rho^{norm}/dp)_{pVCO}$ vs. $T$. The red line represents a fit to the data with $\chi \sim 1/(T - T_{cr})$. The fit gives $T_{cr}$ = -49 $\pm$ 15 K. Error bars on $\chi$ are shown for representative data points. (D) Normalized resistivity $\rho^{norm}$ vs. the generalized distance $h/\theta$.

**Supplementary Figure S9**

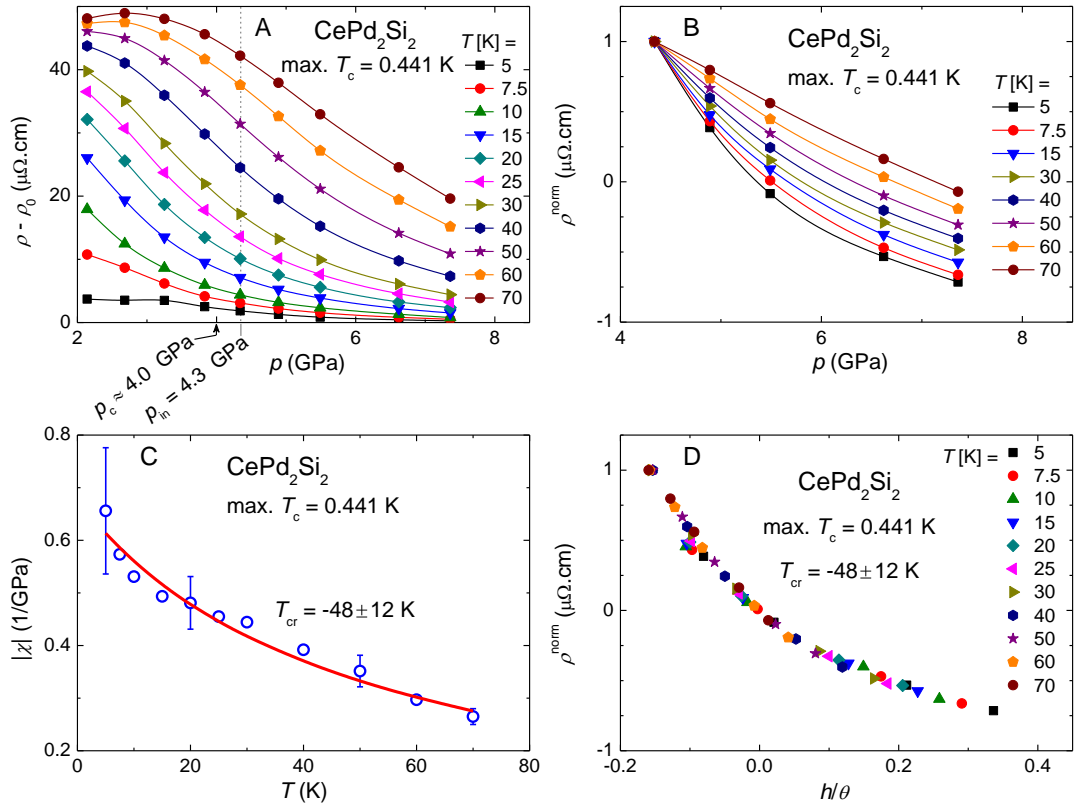

**Fig. S9.** Resistivity scaling analysis of a CePd$_2$Si$_2$ sample with $T_c$ = 0.441 K. $\rho(T)$-data first published in [19]. (A) Resistivity isotherms $\rho^* = \rho - \rho_0$ vs. pressure $p$ at temperatures $T$ from 5 up to 70 K. (B) Normalized resistivity $\rho^{norm}$ vs. $p$. (C) Slope $\chi = (d\rho^{norm}/dp)_{pVCO}$ vs. $T$. The red line represents a fit to the data with $\chi \sim 1/(T - T_{cr})$. The fit gives $T_{cr}$ = -48 $\pm$ 12 K. Error bars on $\chi$ are shown for representative data points. (D) Normalized resistivity $\rho^{norm}$ vs. the generalized distance $h/\theta$.

**Supplementary Figure S10**

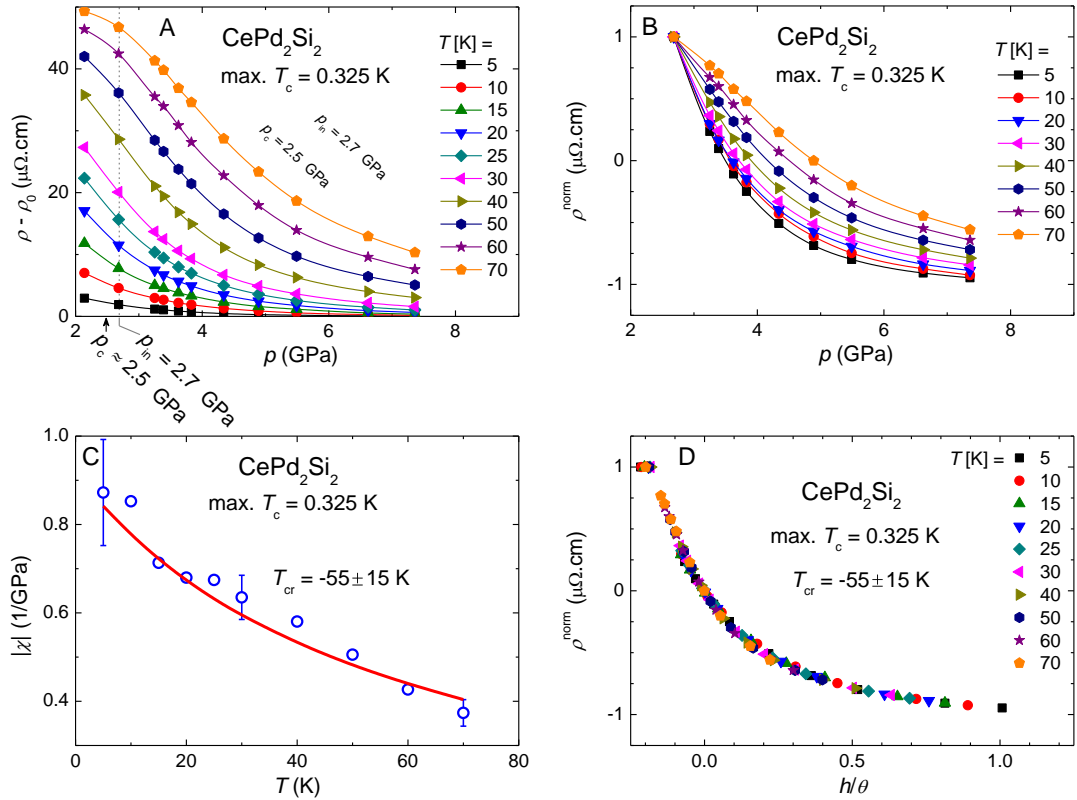

**Fig. S10.** Resistivity scaling analysis of a CePd$_2$Si$_2$ sample with $T_c$ = 0.325 K. $\rho(T)$-data first published in [19]. (A) Resistivity isotherms $\rho^* = \rho - \rho_0$ vs. pressure $p$ at temperatures $T$ from 5 up to 70 K. (B) Normalized resistivity $\rho^{norm}$ vs. $p$. (C) Slope $\chi = (d\rho^{norm}/dp)_{pVCO}$ vs. $T$. The red line represents a fit to the data with $\chi \sim 1/(T - T_{cr})$. The fit gives $T_{cr}$ = -55 $\pm$ 15 K. Error bars on $\chi$ are shown for representative data points. (D) Normalized resistivity $\rho^{norm}$ vs. the generalized distance $h/\theta$.

**Supplementary Figure S11**

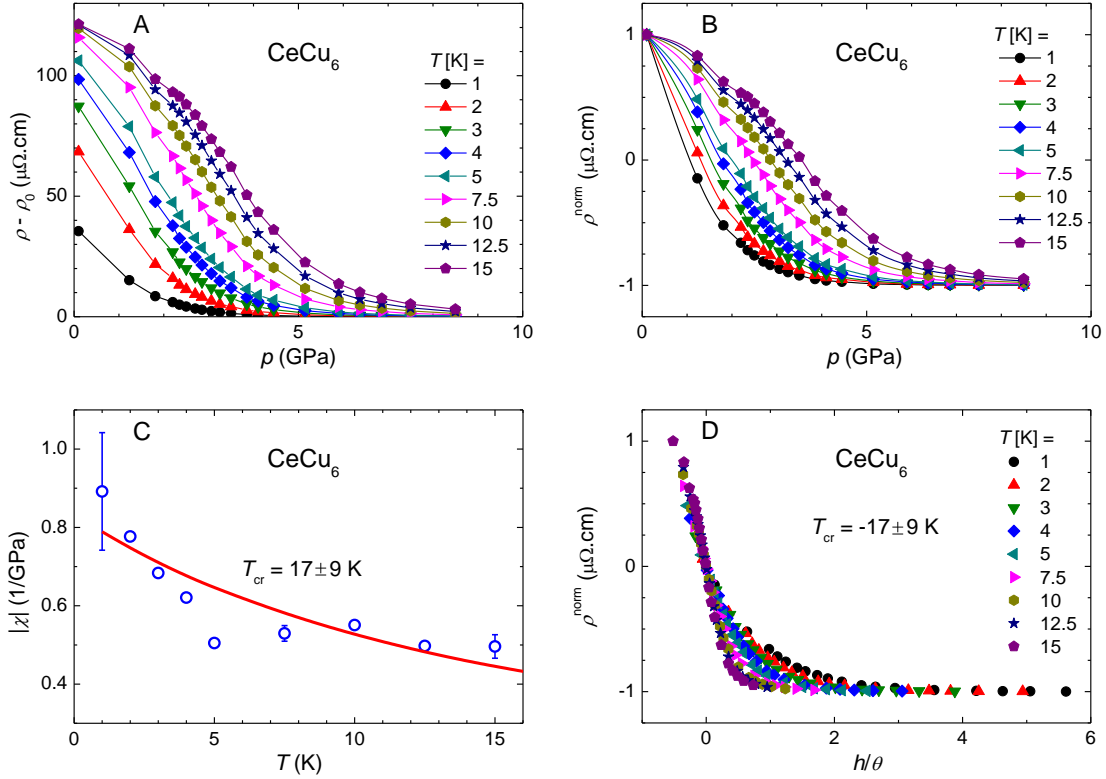

**Fig. S11.** Resistivity scaling analysis of a CeCu$_6$ sample. $\rho(T)$-data obtained in steatite $p$-medium first published in [34]. (A) Resistivity isotherms $\rho^* = \rho - \rho_0$ vs. pressure $p$ at temperatures $T$ from 1 to 15 K. Looking on the "raw" resistivity data $\rho^*$, we suspect that measurement errors for pressures up to 2 GPa are most likely due to the combined effect of error on $p$ and bad thermalization of the sample (fast cooldown). (B) Normalized resistivity $\rho^{\mathrm{norm}}$ vs $p$. (C) Slope $\chi = (\mathrm{d}\rho^{\mathrm{norm}}/\mathrm{d}p)_{p\mathrm{VCO}}$ vs. $T$. The red line represents a fit to the data with $\chi \sim 1/(T - T_{\mathrm{cr}})$. The fit gives $T_{\mathrm{cr}} = $ -17 $\pm$ 9 K. Error bars on $\chi$ are shown for representative data points. The bad overlap of fit and data is presumably due to the measurement errors noted before. Also, a low pressure-run density at the beginning of the pressurization results in a high error on $\chi$. (D) Normalized resistivity $\rho^{\mathrm{norm}}$ vs. the generalized distance $h/\theta$. The curves do not collapse on a single scaling function. For now, it is not clear if this is intrinsic to CeCu$_6$ or due to the measurement errors only. A change of regime ascribed to the crystal field effect [34] may interfere.

**Supplementary Figure S12**

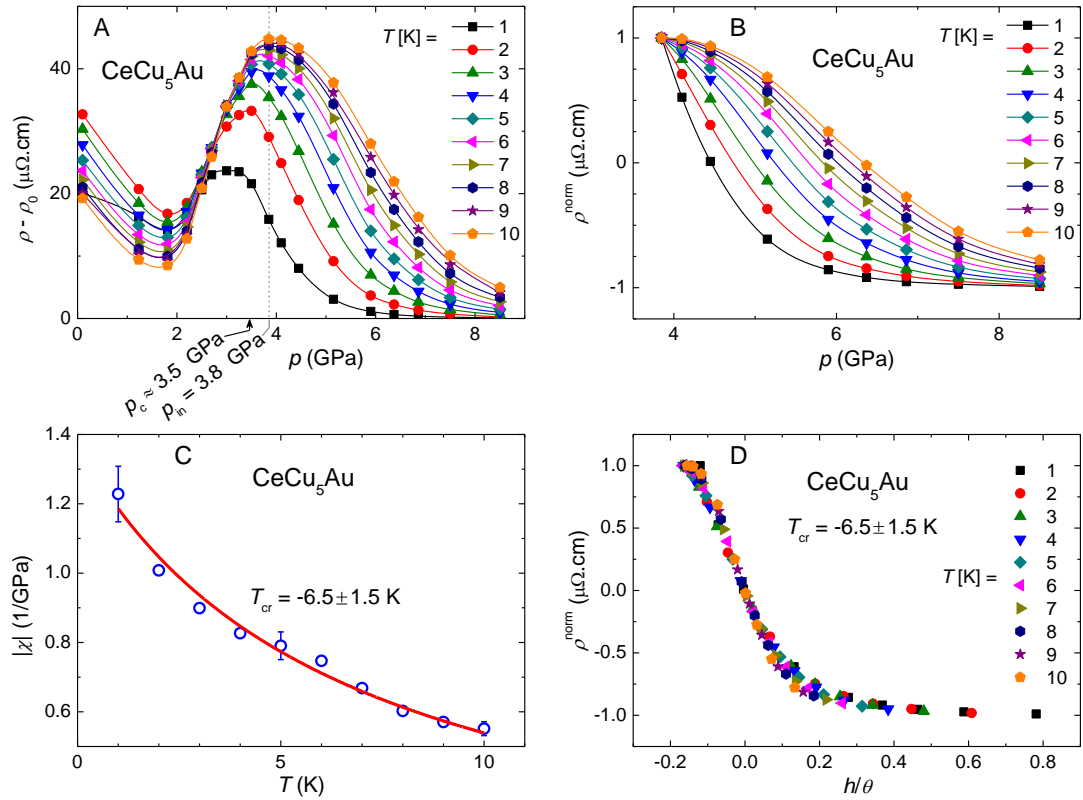

**Fig. S12.** Resistivity scaling analysis of a CeCu$_5$Au sample. $\rho(T)$-data first published in [75]. (A) Resistivity isotherms $\rho^* = \rho - \rho_0$ vs. pressure $p$ at temperatures $T$ from 1 up to 10 K. Note that the scaling analysis is limited to T ≤ 10 K because of a low $T_1{}^{max}$ ~ 25 K at the VCO. (B) Normalized resistivity $\rho^{norm}$ vs. $p$. (C) Slope $\chi = (d\rho^{norm}/dp)_{pVCO}$ vs. $T$. The red line represents a fit to the data with $\chi \sim 1/(T - T_{cr})$. The fit gives $T_{cr}$ = -6.5 ± 1.5 K. Error bars on $\chi$ are shown for representative data points. (D) Normalized resistivity $\rho^{norm}$ vs. the generalized distance $h/\theta$.

**Supplementary Figure S13**

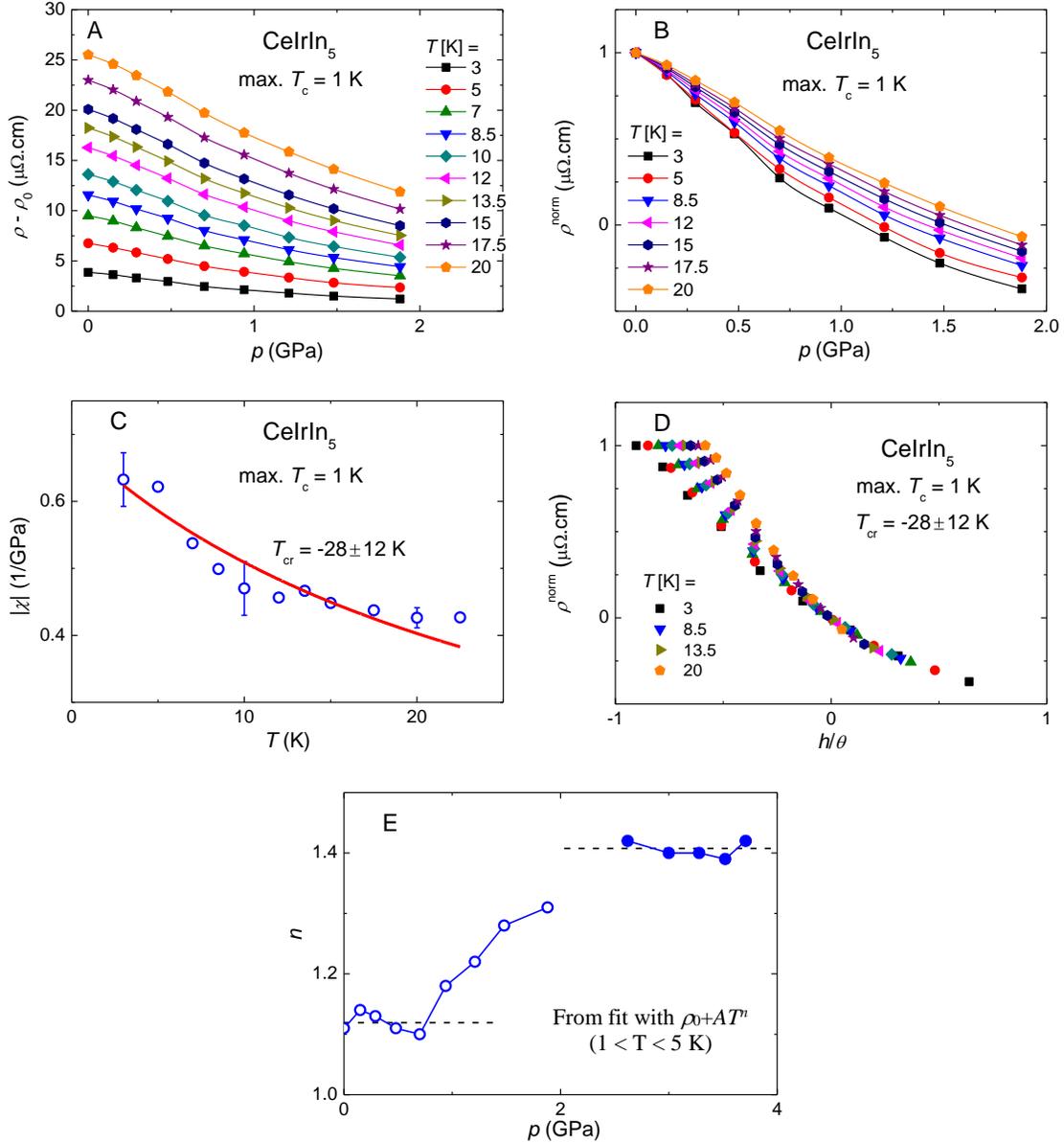

**Fig. S13.** Resistivity scaling analysis of a CeIrIn$_5$ sample. (A) Resistivity isotherms $\rho^* = \rho - \rho_0$ vs. pressure $p$ at temperatures $T$ from 3 up to 20 K. (B) Normalized resistivity $\rho^{norm}$ vs. $p$. (C) Slope $\chi = (d\rho^{norm}/dp)_{pVCO}$ vs. $T$. The red line represents a fit to the data with $\chi \sim 1/(T - T_{cr})$. The fit gives $T_{cr} = -28 \pm 12$ K. Error bars on $\chi$ are shown for representative data points. (D) Normalized resistivity $\rho^{norm}$ vs. the generalized distance $h/\theta$. The $\rho^*$ isotherms collapse on a single curve only for $h/\theta > 0$. (E) Exponent $n$ versus pressure $p$, obtained by fitting low-temperature resistivity $\rho(T)$ with $\rho_0 + AT^n$. Open symbols: data corresponding to Fig. S13A, full symbols: preliminary data from a new high-pressure experiment. The behavior of $n(p)$ hints to a pressure-induced change of regime, which may explain why the isotherms do not collapse for $h/\theta < 0$.

**Supplementary Figure S14**

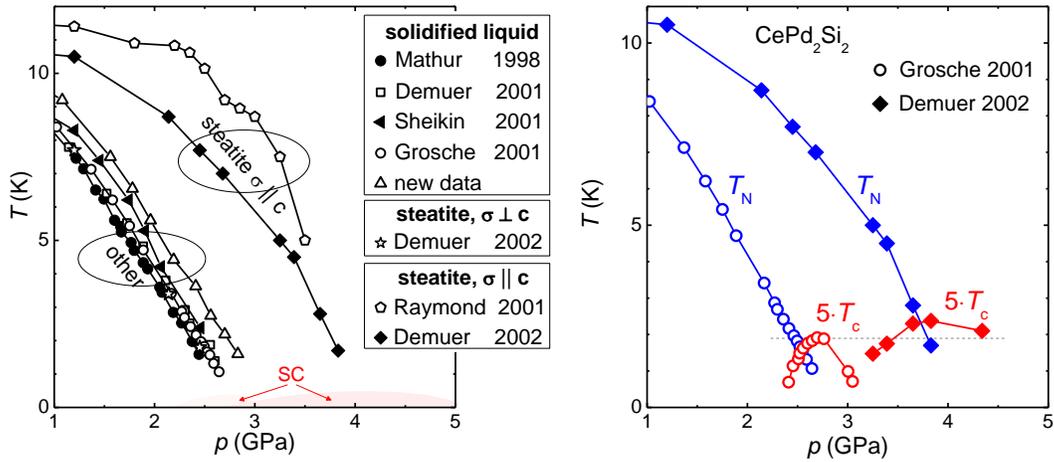

**Fig. S14.** Pressure-temperature phase diagram of CePd$_2$Si$_2$ samples measured in "solidified-liquid" [3,43,54,55] and "soft-solid-steatite" [19,56] pressure-transmitting medium ("solidified-liquid" = He, pentane, Daphne oil). Left-panel: all known data of the $T_N$ transition line. A linear decrease of $T_N$ occurs in "solidified-liquid" or steatite with stress $\boldsymbol{\sigma} \perp \mathbf{c}$. A more rapid vanishing of $T_N$ occurs in steatite with stress $\boldsymbol{\sigma} \parallel \mathbf{c}$. The corresponding superconducting phases are represented schematically. Right panel: Phase diagram of two samples of similar quality as indicated by their residual resistivity $\rho_0 \approx 1.1$ $\mu\Omega$cm. A steeper collapse of the antiferromagnetic transition temperature $T_N$ seems to favor higher superconducting $T_c$. Noteworthy, Demuer *et al.* [19] have observed a strain-driven enhancement of $T_c$. Two samples have been measured simultaneously in steatite with $\boldsymbol{\sigma} \perp \mathbf{c}$ and $\boldsymbol{\sigma} \parallel \mathbf{c}$, respectively. Despite lower sample quality, a higher $T_c$ was found for $\boldsymbol{\sigma} \parallel \mathbf{c}$.